\documentclass[preprint2,trackchanges]{aastex631} 

\usepackage{graphicx}
\usepackage[suffix=]{epstopdf}
\usepackage{natbib}
\usepackage{amsmath}
\usepackage{url}
\usepackage{xspace}
\usepackage{listings}
\usepackage{hyperref}

\providecommand{\e}[1]{\ensuremath{\times 10^{#1}}}

\shorttitle{Technosignatures with Alerts}
\shortauthors{Gallay et al.}

\graphicspath{{./}{figures/}}

\begin{document}

\title{Technosignature Searches with Real-time Alert Brokers}


\author[0009-0003-8543-2258]{Eleanor M. Gallay}
\affiliation{Smith College, Northampton, MA 01063, USA}

\author[0000-0002-0637-835X]{James R. A. Davenport}
\affiliation{Department of Astronomy, University of Washington, Box 351580, Seattle, WA 98195, USA}

\author[0000-0003-4823-129X]{Steve Croft}
\affiliation{Breakthrough Listen, University of Oxford, Department of Physics, Denys Wilkinson Building, Keble Road, Oxford, OX1 3RH, UK}
\affiliation{University of California, Berkeley, 501 Campbell Hall \#3411, Berkeley, CA 94720, USA}
\affiliation{SETI Institute, 339 Bernardo Ave, Suite 200, Mountain View, CA 94043, USA}

\begin{abstract}

We present an exploration of technosignature research that is possible using real-time alert brokers from surveys such as the Zwicky Transient Facility (ZTF) and the upcoming Legacy Survey of Space and Time (LSST). Nine alert brokers currently stream up to 1 million alerts each night from ZTF, and LSST is projected to increase this volume by an order of magnitude. While these brokers are primarily designed to facilitate real-time follow-up of explosive transients such as supernovae, they offer a unique platform to discover rare forms of variability from nearby stars in real time, which is crucial for follow-up and characterization. 
We evaluate the capability for both spatial and temporal searches for extraterrestrial intelligence (SETI) methods using the currently available brokers, and present examples of technosignature searches using ZTF alert and archival data. We have deployed optical SETI techniques, such as planetary transit zone geometries and the SETI Ellipsoid. We have also developed a search for novel high-amplitude stellar dippers, and present a workflow that integrates features available directly through the brokers, as well as post-processing steps that build on the existing capabilities. Though the SETI methods that alert brokers can execute are currently limited, we provide suggestions that may enhance future technosignature and anomaly searches in the era of the Vera C. Rubin Observatory.

\end{abstract}


\section{Introduction} \label{sec:intro}

Modern optical surveys like the Zwicky Transient Facility \citep[][ZTF]{bellm2019} and the upcoming Legacy Survey of Space and Time (LSST) on the Vera C. Rubin Observatory \citep{lsst} push time-domain astronomy towards an age of big data. With over 1 million alerts coming in each night from ZTF, nine different community brokers worldwide enable the public to immediately access transient data. 
The total LSST imaging data is expected to reach $\sim$20 TB per night, far more than can be analyzed by normal methods, necessitating the use of these community brokers for early time domain science.
Seven of these brokers have been selected to receive the full stream of alerts from LSST once the telescope becomes operational in 2025. LSST is projected to increase the volume of transient detection by orders of magnitude, with roughly 10 million alerts expected each night. 
Alert brokers provide researchers and amateur astronomers alike with a means of efficiently classifying alerts and narrowing the data stream to find signals of interest, as well as explore the rapidly growing archive of variability events. 

Across the nine brokers currently streaming ZTF data, features available to the public include filters, alert classification schemes, watchlists of selected targets to monitor, and spatial constraints for where to search for alerts. Filters allow users to narrow the stream of alerts to find a subset of objects that satisfy criteria of interest. Filters are written in Structured Query Language (SQL) and may include any number of conditions given the broker's schema. An example filter is shown in Appendix \ref{ap:SQL}.

Brokers offer a variety of cross-matching and classification systems, which assign each alert to a class of objects such as supernovae or variable stars. The Lasair broker \citep{smith_lasair_2019} uses the Sherlock classification system, which comprises seven classes. Sherlock assigns classifications to transients based on characteristics of the closest catalog match from a number of all-sky surveys and smaller source-specific catalogs. With relativeley few classes and only one identifier for stellar sources, Sherlock caters more toward explosive transients than subtle stellar variability. The ALeRCE broker \citep{forster2021} offers a wider range of classifications, making it well-suited for variable star analysis in addition to transient detection. ALeRCE's pipeline includes a light curve classifier that employs four models and distinguishes transients from stochastic and periodic objects. The classifier assigns every alert a probability score for each of 15 classes, ranging from eclipsing binaries to type Ia supernovae. Users can access the light curve classifier on ALeRCE's website and through their Application Programming Interface (API). While classification schemes vary significantly across broker pipelines, most brokers provide a similar crossmatching feature. Alerts can be crossmatched against several different catalogs, including Gaia \citep{gaia_collaboration_gaia_2021}, 2MASS \citep{skrutskie_two_2006}, and SDSS \citep{york2000}, to identify whether they are coming from a known source. These crossmatches can help enable spatial technosignature searches if the matched sources have well defined parallaxes, as in the Gaia survey. 

Another consistent feature across brokers is a search engine for individual alerts from the stream. Typical search parameters include magnitude limits, Julian dates, and position on the sky. For instance, users can employ a cone search to specify the right ascension, declination, and radius within which to look for alerts.

In addition to cone searches, the Lasair broker offers a unique watchmap feature, enabling users to specify an arbitrary region of sky in which to search for alerts. To create a watchmap, users must upload a region file formatted as a Multi-Order Coverage (MOC) map \citep{fernique_moc_2014}. The process of making a MOC file involves specifying a desired region of sky in standard coordinates such as the International Celestial Reference System (ICRS) and generating a composition of geometrical shapes to optimally fill the region. Lasair then matches alerts against the saved watchmap if they fall inside the portion of the sky it defines. Lasair integrates its filter and watchmap features so that users can query for alerts with watchmap matches. Once saved, filters run constantly, matching alerts with filter criteria as data streams in each night of observation. Users can opt to receive notifications via daily email or Kafka streams of new alerts that pass filters \citep{patterson2019}. 

Users can upload and run public or private filters, watchlists, and watchmaps directly on the Lasair website. Watchlists allow users to monitor up to a few thousand sources for alerts, without having to pull information from their API directly. The Lasair API offers more complex analytical tools than those available on the broker interface, though the filter, watchlist, and watchmap features housed internally enable a range of transient and variable star research with any laptop connected to the broker cloud.

While brokers are all designed with the same purpose of connecting the public with the alert stream, they function in distinctive ways, each suited to different projects and research topics. Watchmaps make Lasair advantageous for spatial and spatiotemporal-driven research, including for technosignature searches, thus we primarily focus on the capacity for anomaly detection with Lasair in this paper.

To distinguish technosignature searches from broader anomaly detection methods, we focus on spatial and temporal constraints that pertain specifically to preexisting SETI techniques. It is impossible to know what a technosignature looks like, where it will come from, or when it will arrive. We must therefore make logical guesses about what to look for, where to look, and when to look, to find a signal of interest. In radio SETI, the typical answer of what to look for is a narrow-band Doppler drifted signal, as in \citet{choza_breakthrough_2024}. Recent work has also employed machine learning techniques to reject false-positive technosignature candidates due to radiofrequency interference \citep{ma_deep-learning_2023}. This method rejects RFI signals based on their morphology, as we have a clear sense of what not to look for. There are many approaches to deciding where to look, such as identifying specific targets to optimize the chances of observing a faint radio signal \citep{isaacson_breakthrough_2017} or catching a transmission between exoplanets \citep{tusay_radio_2024, siemion_11-19_2013}. Other spatially defined search methods include looking near the galactic center \citep{gajjar_breakthrough_2021} or within the Earth Transit Zone \citep{sheikh_breakthrough_2020}. The SETI Ellipsoid \citep{lemarchand1994, davenport2022} and Seto \citep{seto2021} schemes attempt to answer the question of when to look, anticipating coordinated singles with significant astrophysical events. All three questions, what, where, and when, attempt to discern interesting signals from simply unusual ones. In this paper, we use the Lasair broker to employ techniques like Earth Transit Zone searches, SETI Ellipsoid and Seto schemes, and searches for novel stellar dippers like Boyajian's star \citep{boyajian_planet_2016} and ASASSN-21qj \citep{rizzo-smith2021}. We emphasize that these search methods may return signals with SETI implications, but may also discover anomalies with astrophysical origins.

\begin{figure*}[!t]
    \centering
    \includegraphics[width=6in]{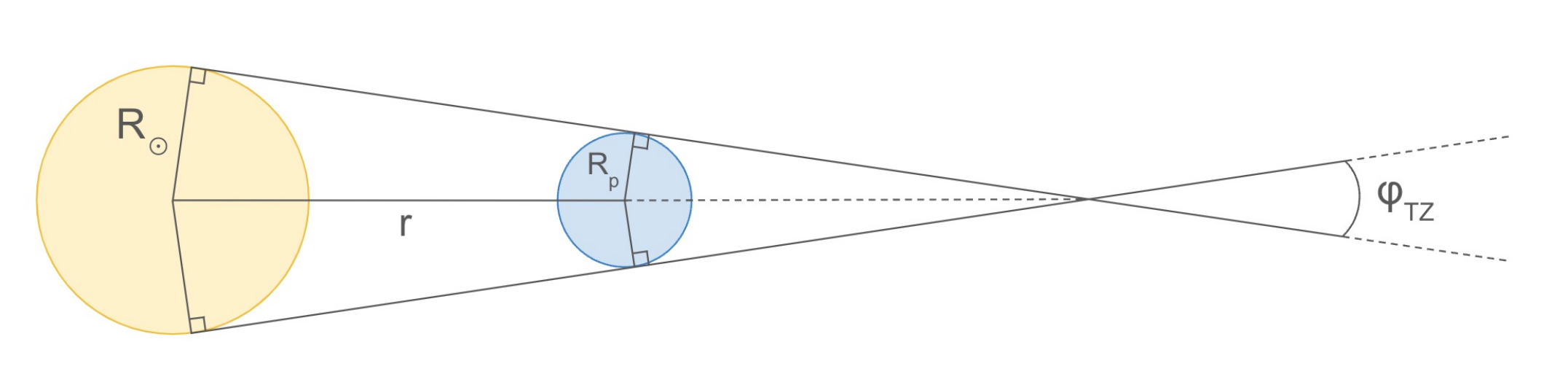}
    \caption{Transit zone angle geometry. The yellow and blue circles represent the Sun and planet, respectively (sizes not drawn to scale). $R_{\odot}$ is the solar radius and $R_{\text{p}}$ is the planet's radius. The average distance between the planet and the Sun is r (1 AU for Earth), and $\phi_{\text{TZ}}$ is the full transit zone angle. The transit zone is a projection of $\phi_{\text{TZ}}$ onto the sky. Any theoretical sensor in this strip of sky could detect the planet's existence via the transit method. Using similarity reasoning, we calculate $\phi_{\text{TZ}}$ = $2\arcsin((R_{\odot}-R_{p})/r)$.}
    \label{fig:triangles}
\end{figure*}

While their main focus is to report new supernovae and explosive transients, brokers also offer a unique opportunity for optical technosignature searches in real time. Brokers enable searches for spatial and temporal signals. This paper discusses examples of the types of SETI projects that are possible with the tools currently offered by alert brokers. We present a workflow that integrates features available directly through the Lasair broker with post-processing steps that build on the existing broker capabilities. We examine the extent to which brokers can be used to conduct SETI research and offer suggestions for ways in which brokers can be modified to facilitate technosignature searches along with anomaly detection more broadly. 

\section{Spatial Signals}

\begin{figure*}[!ht]
    \centering
    \includegraphics[width=0.5\linewidth]{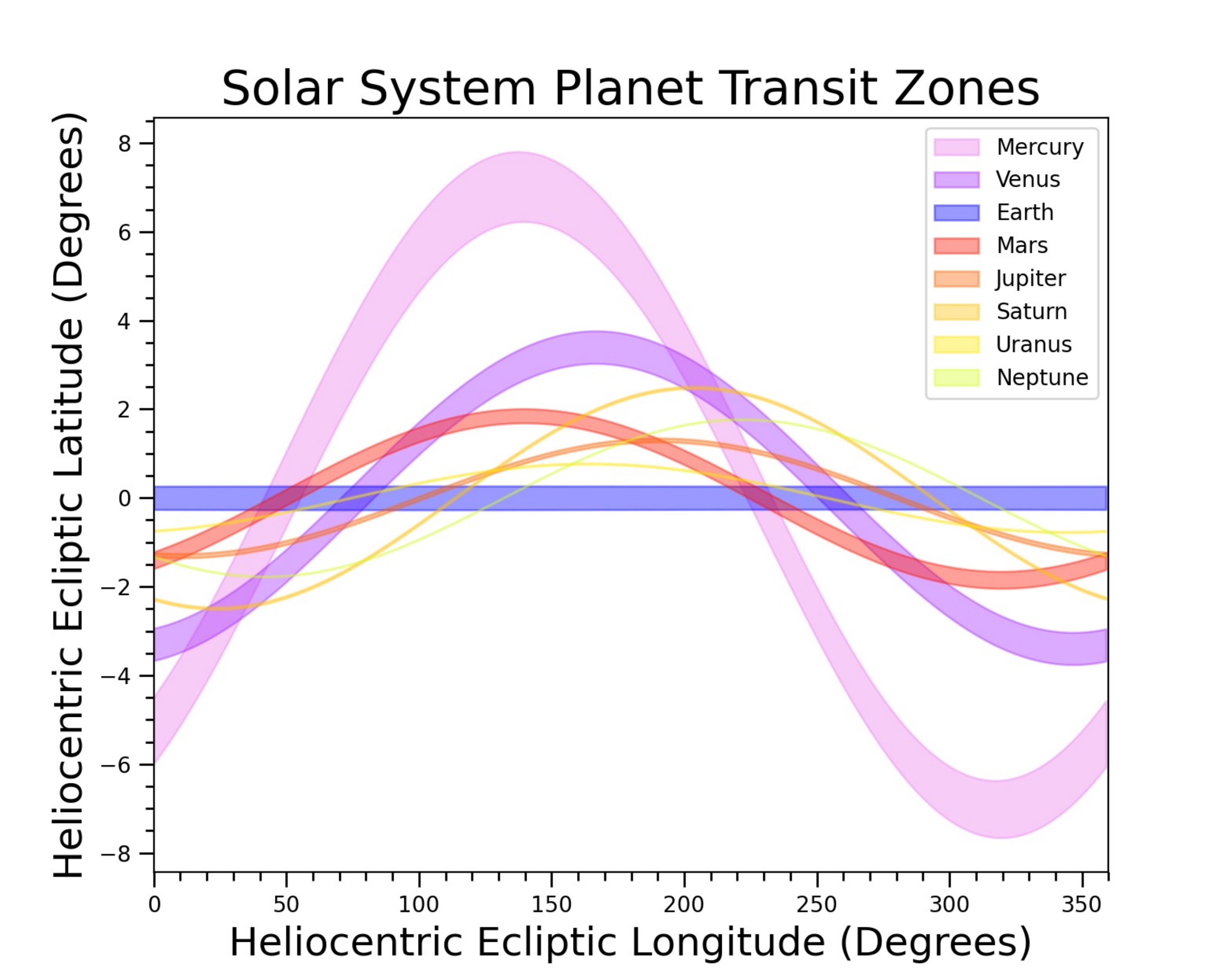}
    \includegraphics[width=0.49\linewidth]{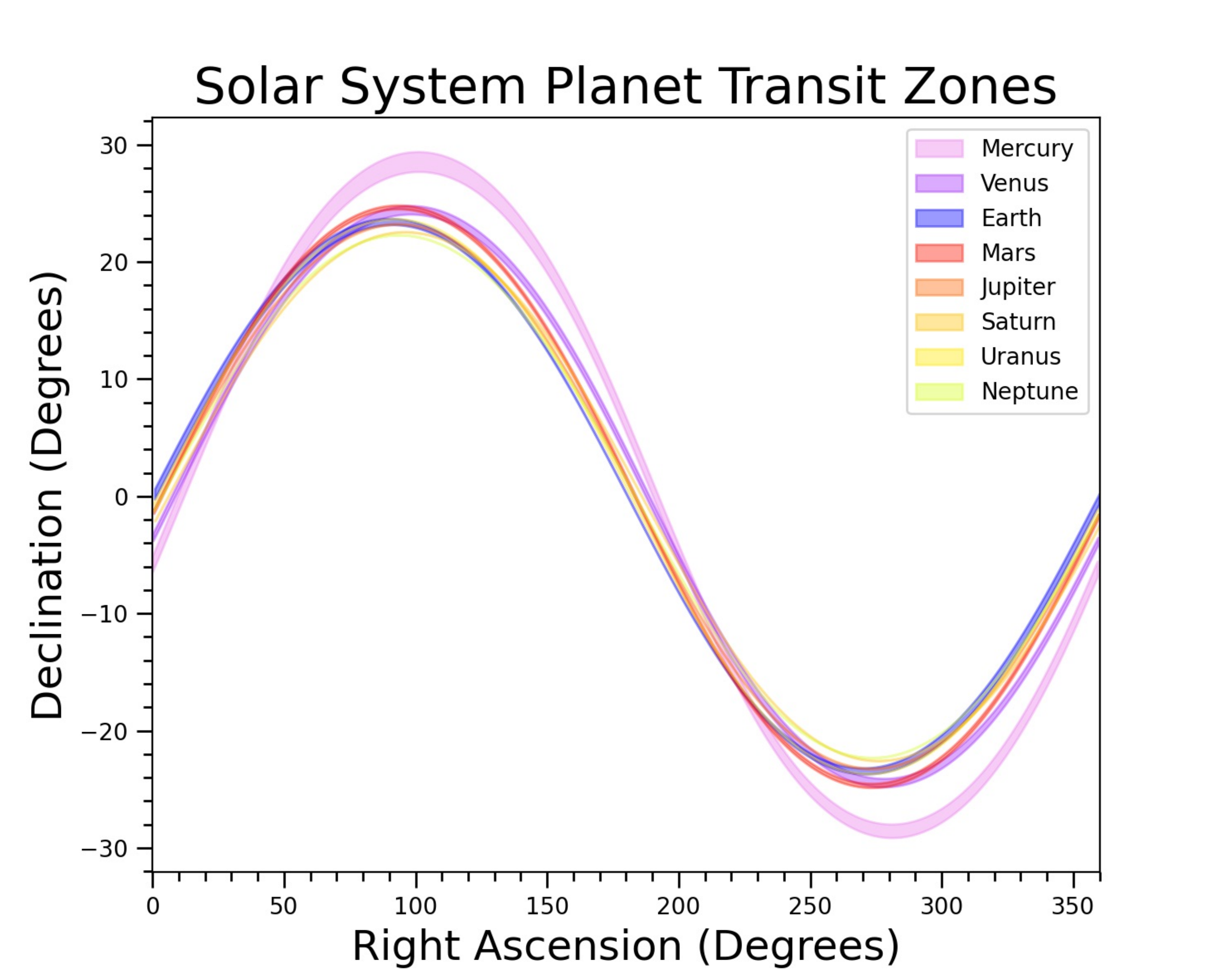}
    \caption{
    Transit zones for all solar system planets are plotted in heliocentric ecliptic coordinates (left) and in celestial coordinates (right). Each planet's transit zone is shown as a colored sinusoidal band. Sensors located in the colored regions of sky could detect the corresponding planets as they transit the Sun. Overlaps between colored bands indicate regions where multiple planets could be detected. Note there are 28 regions of overlap between sets of two planet transit zones and 8 regions of overlap for sets of three planets. We adapt code from \citet{wells_transit_2018} and use the updated transit zone geometry (Figure \ref{fig:triangles}) to calculate and plot the full transit zone angles as $\phi_{\text{TZ}}$ = $2\arcsin((R_{\odot}-R_{p})/r)$.}
    \label{fig:zones}
\end{figure*}

\begin{figure*}[!ht]
    \centering
    \includegraphics[width=0.5\linewidth]{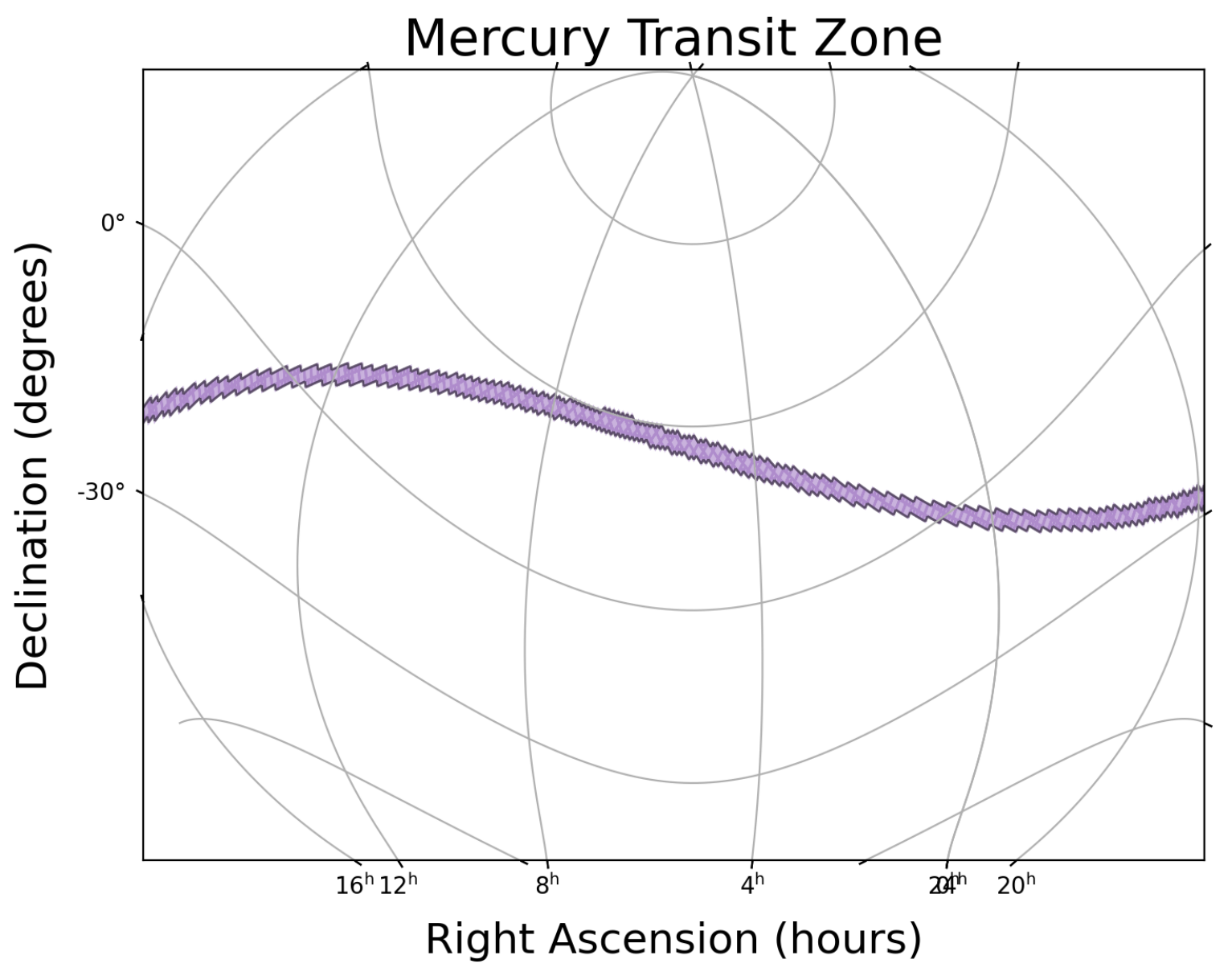}
    \includegraphics[width=0.49\linewidth]{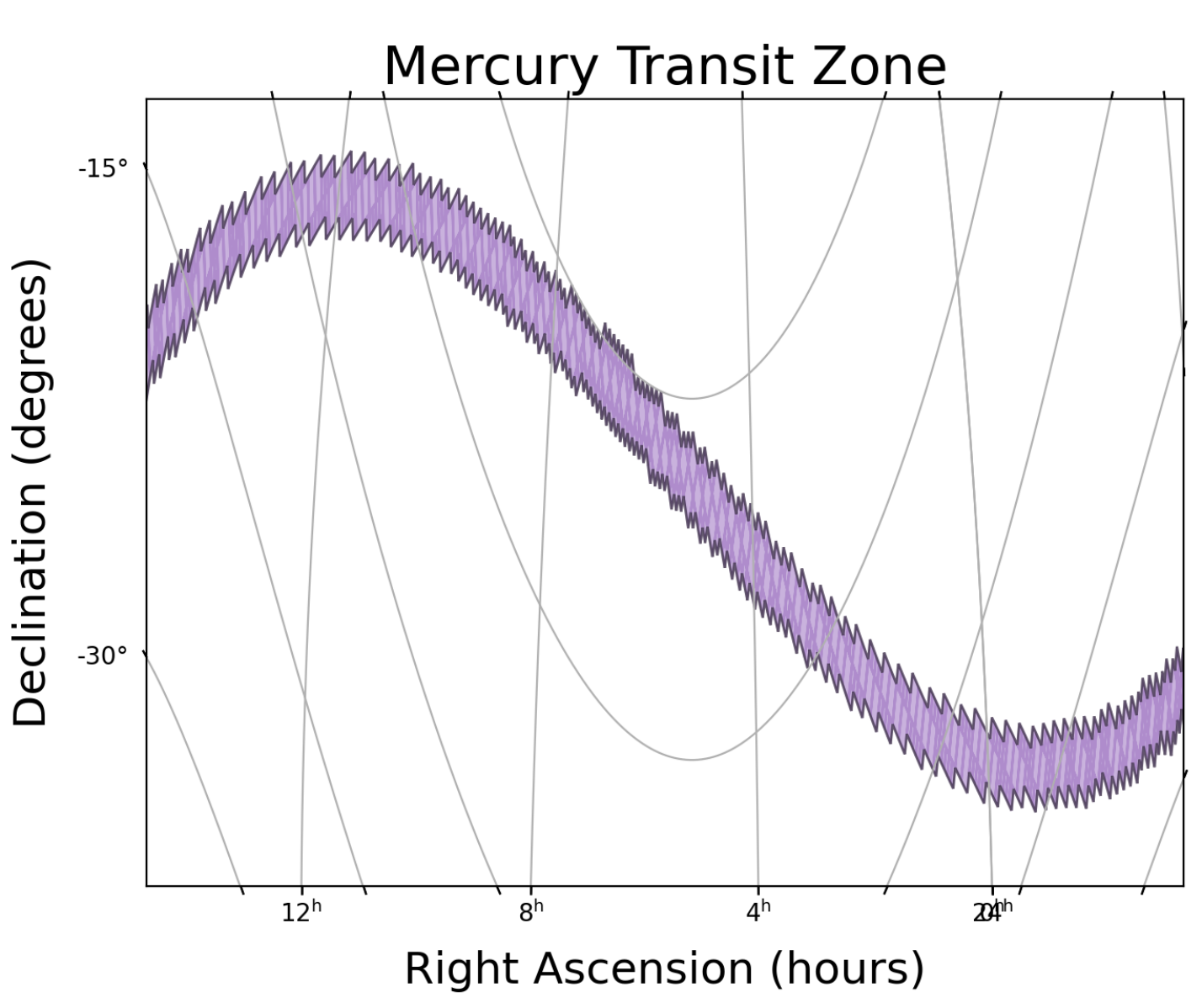}
    \caption{
    Multi-Order Coverage map for the Mercury Transit Zone is plotted in purple with a HEALPix cell resolution of 12. The plot is a projection of the world coordinate system (WCS) to the pixel image coordinate system. The right-hand plot more clearly shows the jagged edge of the MOC rendering, which slightly overestimates the total transit zone region.}
    \label{fig:mercury}
\end{figure*}

Lasair's watchmap and filter features enable spatial technosignature searches directly in the broker. We present an example of this kind of search using the Earth Transit Zone along with all planetary transit zones in the solar system. The Earth Transit Zone (ETZ), as defined by \citet{heller2016}, is the region of the sky from which Earth can be seen as a transiting exoplanet. Signals found within the ETZ may be signs of attempted communication or broadcasting to Earth, as a theoretical extraterrestrial civilization could become aware of the planet's existence via transit detection. This logic motivates searches for signals in all planetary transit zones to optimize the chance of technosignature detection because sensors in these regions could know of the existence of any solar system planets. 

We calculate the transit zone angles for the solar system planets using the geometry shown in Figure \ref{fig:triangles}. The full transit zone angle encompasses the range of sky where the entire planet occults the Sun, while the grazing transit angle is the region where any fraction of the planet occults the Sun. We make a slight modification to the full transit zone calculations done by \citet{wells_transit_2018} and \citet{heller_search_2016} by defining the full transit zone angle $\phi$ as twice the angle between the line intersecting the solar and planetary centers and the line between the tangent points of the two spheres. We calculate this angle as $\phi_{\text{TZ}}$ = $2\arcsin((R_{\odot}-R_{p})/r)$ using properties of similar triangles. For our planetary transit zone modeling, it is important to note that when using the small-angle approximation, our full transit zone angle calculation and that of \citet{heller_search_2016} reduce to the same result of $\phi_{\text{TZ}}$ $\approx$ $2R_{\odot}/r$, as used by \citet{castellano_visibility_2004}.

Following the approach used by \citet{wells_transit_2018} we use data from the NASA planetary fact sheets and JPL Horizons to calculate the planetary transit zones. We adapt code from \citet{wells_transit_2018} to plot all eight transit zones in heliocentric ecliptic coordinates (Figure \ref{fig:zones}, left). These transit zone calculations and plots are derived from the corrected geometry shown in Figure \ref{fig:triangles}. We transform from ecliptic to celestial coordinates (Figure \ref{fig:zones}, right) and generate a MOC file for each transit zone with HEALPix level 12, corresponding to a cell resolution of up to 51.53\arcsec
\citep{healpix}. 
Figure \ref{fig:mercury} shows a plot of the MOC map for the Mercury Transit Zone.  After saving the MOC files for all planetary transit zones, we uploaded them to Lasair as watchmaps, where they are now running continuously and are publicly available. 

We used these transit zone watchmaps to create 44 public filters on Lasair. The filters look for stars cross-matched with the Gaia catalog that flag alerts within each planetary transit zone and each region where the zones overlap. There are 28 overlap regions between sets of two transit zones and eight overlap regions between sets of three transit zones. Transit zone filters could be further refined by restricting alert or object characteristics to search for a specific type of signal or technosignature.

The techniques used to search the planetary transit zones for alerts can be adapted for a range of other spatially constrained anomaly searches, including SETI research in any geometrically defined region, such as the galactic plane or bulge. The watchmap feature makes the Lasair broker advantageous for mapping and monitoring regions of interest for SETI candidates.

\section{Temporal Signals}

\begin{figure*}[!ht]
    \centering
    \includegraphics[width=5in]{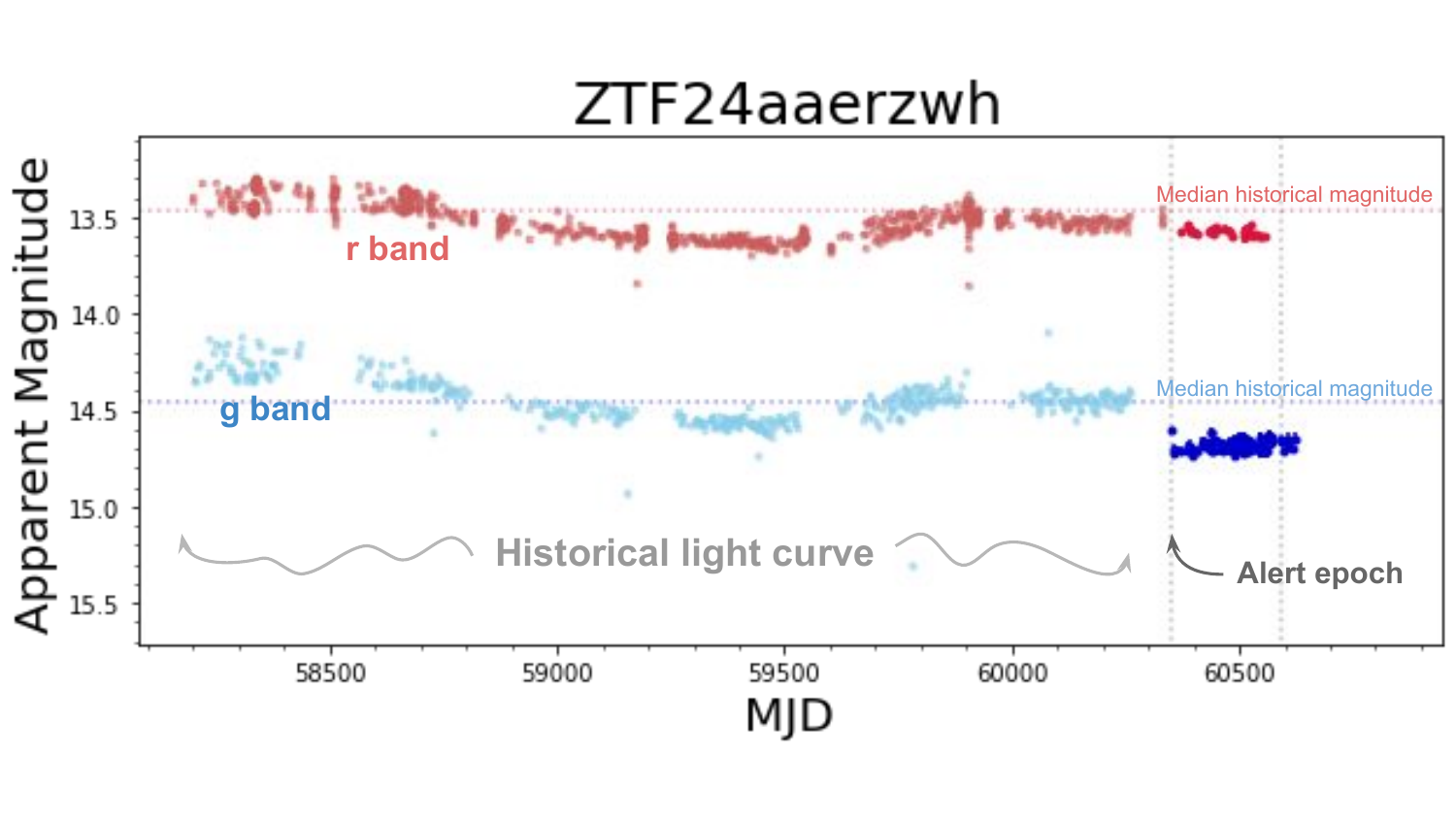}
    \caption{Light curve for ZTF24aaerzwh. This source was crossed-matched by Lasair with 4479767064322208640, a G=13.40 mag stellar source found in the GAIA/PS1 catalogs. The source has been classified in the Zwicky Transient Facility catalog of periodic variable stars as a BY Draconis variable. The data plotted is an amalgamation of historical data pulled from the ZTF archive and alert packet data pulled with the Lasair API query. R-band and G-band data are shown in red and blue, respectively. Data pulled from the ZTF alert packet is plotted in a darker shade. This alert packet data is also distinguished from archival data because it lies to the right of the alert epoch line, indicating the object's discovery date in Lasair. The right-most dotted line indicates the `MJD min' or thirty days before the object's most recent alert.}
    \label{fig:lpv}
\end{figure*}

The time-series data contained in ZTF alert packets and the transient archive enable searches for anomalous and novel stellar variability. These data include brightness observations across up to three filters, along with Julian dates for each flux reading. We can construct light curves from alert packet data to track objects’ variability over time and identify unusual behavior. When evaluating what kinds of temporal signals are well-suited for detection with alert brokers, we must contend with factors inherent to the alert brokers' internal structure and the nature of ground-based observing on optical telescopes. 

The alert system identifies optical transients by subtracting brightness measurements from reference fluxes taken before the survey. The absolute value of each subtraction is called a difference magnitude, and is stored in an object's alert packet in the broker cloud. Transients must be variable enough to trigger an alert, though a minimum amplitude threshold built into the brokers is not publicized. We can rule out searches for subtle oddities in stellar light curves and insist that projects investigating temporal signals with alert brokers must explore more intense fluctuations in brightness. 

Temporal signal searches with alert brokers must contend with typical challenges associated with ground-based observing, such as discontinuous or patchy data due to unpredictable weather conditions. Inconsistent data make it harder to identify subtle patterns in light-curve behavior, unless the time scale of variability is long enough to reveal trends despite missing information. Alert brokers are thus not suitable for projects that require precise measurement of incremental fluctuations over short time scales, such as searches for transit timing variations. However, brokers have a greater chance of picking up long-period variables. In one case, we happened upon an alert that captured an unusual dimming event from ZTF24aaerzwh, a star classified as a BY Draconis variable with rotation periodicity by \citet{chen_zwicky_2020}. This light curve caught our interest when we plotted the alert packet data alongside historical data pulled from the ZTF archive, as shown in Figure \ref{fig:lpv}. Data from ZTF and the Transiting Exoplanet Survey Satellite (TESS) Sector 80 \citep{tess} suggest that this source has a period of around 1 day. In addition to this short-period variability, Figure \ref{fig:lpv} shows a dimming event that spans approximately 1500 days with an amplitude of around 0.5 magnitudes. The initial drop in brightness appears around MJD 58700, after which the source's 1-day periodicity seems to stop. The dimming event is asymmetrical, with a more gradual drop in brightness around MJD 58700 compared to a relatively rapid increase around MJD 59500. The source appears to be a main sequence star, as illustrated in Figure \ref{fig:cmd}, rendering the dip in brightness particularly curious. The cause of the dimming event remains unclear. The broker did not flag this object as an alert until near the end of its long-period variable cycle, before which we infer that the change in brightness was too gradual to pass the minimum amplitude threshold. We anticipate that the broker framework could be used to identify other long-period variables, given a sufficient amplitude. 

Notable phenomena for technosignature research, including light curves with outliers or anomalous behavior, may be ideal candidates for real-time detection with alert brokers. Abrupt fading events like ASASSN-21qj \citep{rizzo-smith2021} show potential for discovery with LSST data and current broker technology. Projects involving considerable light curve analysis may not be possible with the alert brokers alone but can be conducted by implementing user-facing features in conjunction with queries through broker and ZTF APIs.

\section{Spatiotemporal Signals}
One of the unique strengths of wide-field, time domain surveys is the ability to conduct searches for spatiotemporal signals. This includes events that occur at specific locations on the sky and times, such as searching a transit zone at the moment that the Earth's transit shadow intersects a given star. Within the alert broker framework, spatiotemporal anomaly and technosignature searches are facilitated by the standard watchlist and filter features, as well as the watchmaps included in the Lasair broker. 

So far, much of the spatiotemporal technosiganture work has focused on signals that are coordinated in time and space with noteworthy astronomical events. The so-called SETI Ellipsoid approach \citep{lemarchand1994, davenport2022} computes the expected arrival time at Earth from an extraterrestrial agent who has observed a synchronizing event, such as SN 1987A, and broadcast a coordinated signal. This requires monitoring stars over a wide field of view, and with precise distance estimates to accurately constrain the expected signal arrival time at Earth. For current SETI Ellipsoid projects, cross-matching of Alerts to distance catalogs such as Gaia \citep{gaia_edr3} must be available in real time. Other related schemes, such as those proposed by \citet{seto2021}, provide a constantly changing annulus on the sky to search for coordinated signals, which do not require knowing the target stars' distance, and can be easily computed within broker software frameworks.

Following a SETI Ellipsoid approach, such a project could begin with creating a watchlist for objects crossing the 3-dimensional  ellipsoid within a window of time (e.g. 0.1 years, translating to 0.1 lightyear distance around the ellipsoid boundary). These searches have been carried out using Gaia distances for variable stars observed with Gaia \citep{davenport2022,nilipour2023} and the TESS mission \citep{cabrales2024}. Lasair would then compile a list for follow-up analysis of all of Ellipsoid-crossing objects  that have triggered an alert. The Ellipsoid distance can instead be geometrically computed for the entire alert catalog in a given night, provided the stars have known distances that can be filtered over with the broker. This would eliminate the need to redefine a fixed watchlist for Ellipsoid crossing objects every week or month. However, two challenges remain: 1) no broker currently includes distances for all Gaia stars for use in filtering ZTF alerts, and 2) approximately 90\% of stars in the Rubin data stream will be too faint for precise distance constraints from Gaia. We therefore do not expect the SETI Ellipsoid to be the most productive technosignature approach with Rubin alerts, but instead highlight the need for brokers to support potentially complex spatiotemporal filtering, which may reveal important classes of anomalies.

\section{Example workflow: Searching for new stellar dippers}

Using the Lasair broker and data from the public ZTF stream, we have developed a workflow for discovering candidate stars for possible SETI follow-up. Our candidates of interest come from a temporal signal search for new stellar dippers in the ZFT alerts. These objects are stars with a historically constant luminosity that drop suddenly in brightness for a reason unexplained by classical stellar variability or other astrophysical phenomena. Given a sufficiently high amplitude, such anomalous dips may be candidates for SETI follow-up observations to confirm an artificial occultation. 

Our workflow aims to single out alerts that indicate a sudden drop in brightness compared to a historically quiet light curve. We narrow the alert stream in two stages, first filtering with the broker and then imposing additional constraints based on data pulled from historic ZTF lightcurves. A similar workflow can be adapted for optical technosignature searches using data from LSST once the Rubin Observatory becomes operational in 2025.

\subsection{Filtering within the broker}

To access the complete alert packet for a given transient and plot the associated light curve, users must run a SQL query. This query can either be run on the broker interface as a filter or directly through the broker's API. The former requires the filter to be streamed via Kafka so that filter results can be pulled with the broker API. We chose to run the query through the API to avoid this intermediate step, though both methods are possible. 

Queries are limited by the information provided in the brokers' schema. Columns include celestial and galactic coordinates, and difference magnitude statistics for g, r, and i band data. However, the schema lacks some key information such as the date of the object's first alert and object identification in the ZTF archive. The object identification given in Lasair is distinct from the object identification in the ZTF archive. The Lasair identification is uniquely given to an object the first time it enters the Lasair database and remains the same for every subsequent alert. This identification encodes the year of the object's first alert, which we can scrape in our SQL query to limit our search to objects which first flagged alerts in a given year. Precise and complete temporal information would be a valuable addition to the broker schema to facilitate searches for novel and anomalous sources. 

We include our full SQL query for stars that have recently dipped in brightness in Appendix \ref{ap:SQL}. We require that objects have triggered alerts in the past day, such that running the filter each morning will display objects that triggered alerts the night before. Other filter criteria include that the objects first triggered alerts in 2024, have cross-matches with the Gaia catalog, are within 0.5 arcseconds from the best source match from the Sherlock classifier, have at least ten data points in their alert packets, and have closest source matches that are brighter than 16th magnitude in either the g or r band. To filter for disappearing stars, or stars that drop suddenly in brightness, we require that all alert-packet magnitudes be fainter than the source's reference magnitude. 

The query returns a list of objects and their alert packet data. On a typical night with roughly 1 million alerts observed by ZTF, this initial query reduces the stream to under 20 candidates. The vast majority of these remaining sources are not viable candidates for SETI follow-up. We must conduct further analysis beyond the capabilities of the broker in order to advance our search. 

\subsection{Filtering beyond the broker}

After passing the ZTF alert stream through an initial filter, either on the Lasair interface or API, we use information contained in the alert packets of the remaining candidates to plot light curves and impose additional constraints. For a given object, the alert packet contains time series data ranging from thirty days before the alert in question to the most recent ZTF observation. In order to construct the complete light curve for an object, we must convert alert packet data from difference magnitudes to apparent magnitudes and pull historical data from the ZTF archive.

\subsubsection{Apparent Magnitude Calculations}

Difference magnitudes are sufficient to characterize supernovae, but they conceal information essential to the study of variable stars and anomalous dimming events. For instance, Lasair stores difference magnitudes and their respective signs as separate variables in the alert. Furthermore, when searching for anomalies in the context of objects' historical behavior, it is necessary to recover apparent magnitudes in the object's alert epoch in order to compare alert and archival light curves. To obtain apparent magnitudes, we wrote a function that converts the reference and difference magnitudes to fluxes, adds or subtracts depending on the sign of the difference, and then transforms the result back into units of magnitude. We pull our function inputs from columns in the ZTF alert packet, as this information is unavailable in the Lasair schema. The `magnr' column contains the apparent magnitude of the nearest source to the alert in the reference image. The `magpsf' column contains the magnitude of the alert from PSF-fit photometry. This is the difference magnitude displayed by Lasair. Finally, the `isdiffpos' column contains the sign of the difference magnitude. The two possible values for this column are `t' and `f', indicating whether the alert is brighter or fainter, respectively, than the reference image. We also pull uncertainties for the reference and alert measurements to propagate the error when converting from difference to apparent magnitudes. 

Given that historical data from the ZTF archive contains solely apparent magnitudes, we would advocate for the inclusion of apparent magnitudes in the alert packets to facilitate comparative analysis between alert and archival light curves. 

\subsubsection{Archival Queries}

We pull archival light curves for all candidates that pass our SQL query each night. The ztfquery python package \citep{rigault_ztfquery_2018} allows us to plot light curves for regions of the sky within one arcsecond of the candidates' positions. We are unable to pull the historic data for singular objects of interest due to the lack of consistency in object identifiers across ZTF and broker databases. There is not currently a naming convention for alerts among brokers, and neither the Lasair schema nor alert packet contains the archival object identification. For our research purposes, the one-arcsecond search region works well enough to get a general sense of historical patterns. Still, any link between object identifiers, such as including the archival object identification in alert packets, would ensure we are not conflating data from multiple sources. 

Once we have converted alert epoch data to apparent magnitudes and compiled historical data, we plot each candidate's complete light curve, including g, r, and i band data, to the extent that data is available. 

\subsubsection{Statistical Tests}

\begin{figure*}[!ht]
    \centering
    \includegraphics[width=5.5in]{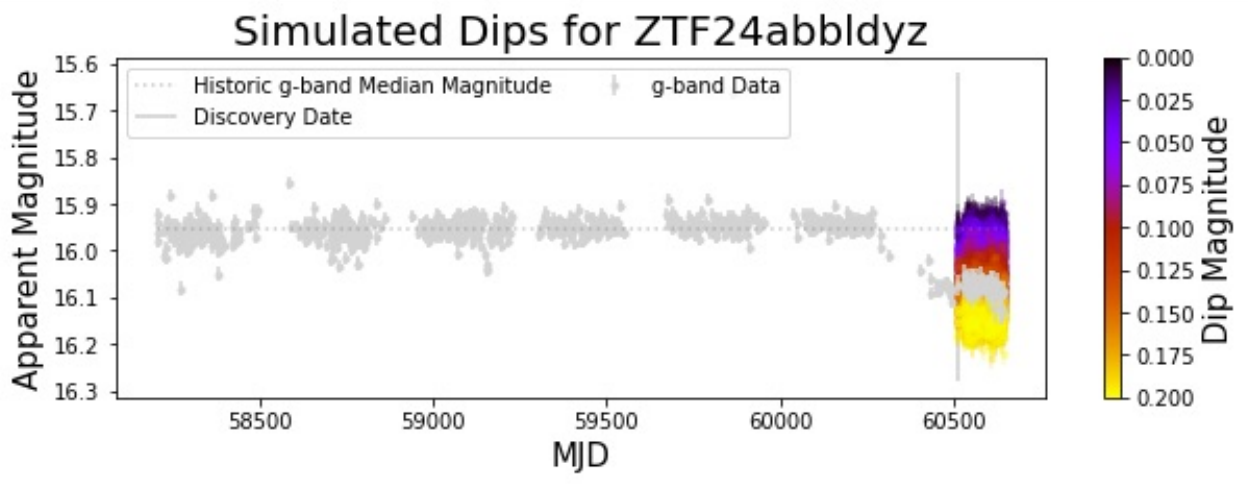}
    \caption{
    Simulated dips of varying amplitudes are plotted along with g-band data from ZTF24abbldyz. Simulated dips are plotted in color, where each color indicates a different amplitude. Real data for ZTF24abbldyz is plotted in greyscale. The horizontal dotted line indicates the median magnitude of historical g-band for ZTF24abbldyz data at least 100 days older than the object's discovery date in Lasair. The solid vertical line marks this discovery date, and grey points plotted to the right of the line are included in the object's ZTF alert packet. We plot simulated dips in 0.01 magnitude increments ranging from 0 to 0.2 magnitudes. We use our simulated data to perform the two-sample K-S test and investigate how p-value depends on dip amplitude.}
    \label{fig:sims}
\end{figure*}

\begin{figure}[!h]
    \centering
    \includegraphics[width=0.9\linewidth]{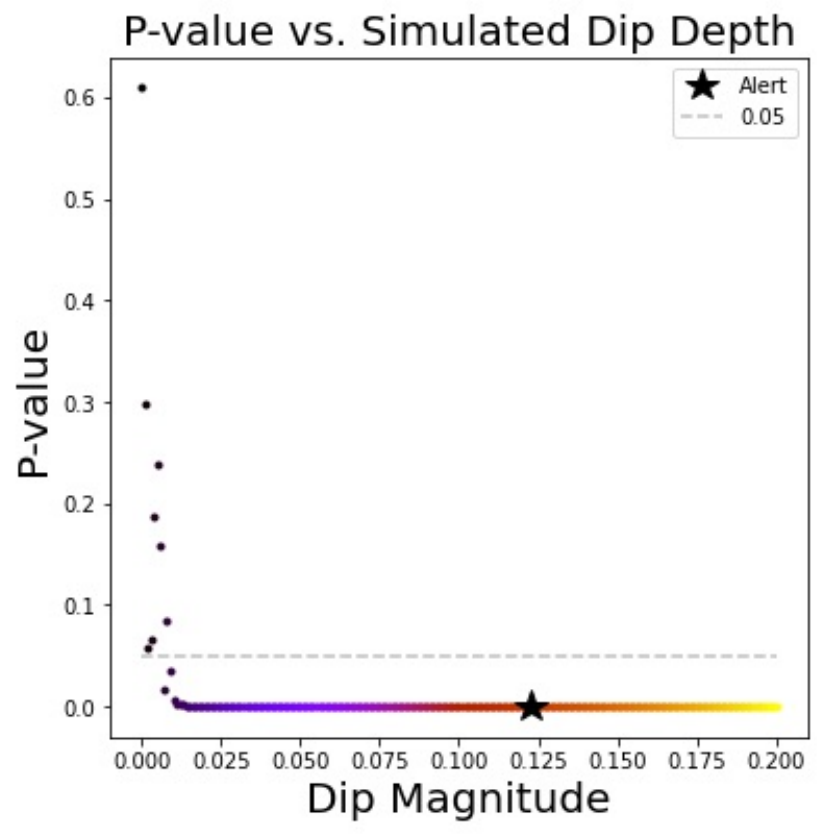}
    \caption{
    P-value from the two-sample K-S test is plotted against simulated dip depth for ZTF24abbldyz, where color corresponds to dip magnitude as in Figure \ref{fig:sims}. The horizontal red line indicates a p-value of 0.05, the cut-off for our two-sample K-S test with a 95\% confidence interval. The p-value for the actual alert data is starred. As expected, p-value decreases with dip amplitude and the relationship appears exponential. The 0.05 p-value cutoff successfully eliminates candidates with dip amplitudes close to zero and recognizes dip amplitudes greater than about 0.01 magnitudes as statistically significant. }
    \label{fig:p-values}
\end{figure}

To narrow our search to include candidates that were historically constant in brightness and whose alerts mark the first instance of a drop in brightness, we perform two statistical tests on the time-series data. These are the chi-squared test and the two-sample Kolmogorov-Smirnov (K-S) test. When performing these tests, we define our historical data sample as observations taken at least 100 days before the source's discovery date and our alert data sample as observations taken at least 30 days after the discovery date. The discovery date is the first data point in a source's alert epoch, marking 30 days before the first data point that flagged an alert. We impose an additional 100-day window between historical and alert data to ensure that if a star begins to dip in brightness before it is flagged as an alert, this dip does not interfere with the variability and distribution of the historical sample. 

The chi-square test measures the variability of the historic ZTF data. We calculate the chi-squared statistic $\chi^2$ as:

\begin{equation}
\chi^2 = \frac{1}{N} \sum_{i=1}^{N} \left( \frac{O_i - E}{\sigma_i} \right)^2.    
\end{equation}

For a given filter, $N$ is the number of data points in the historical sample, $O_i$ is the magnitude measured for the $i$th data point, $\sigma_i$ is the associated error, and $E$ is the median magnitude for the historical sample.

We only keep candidates with at least 100 historical data points and chi-square statistics less than 15 in every filter in which data was taken. To impose a second test of historical variability, we calculate the 5th to 95th percentile range of the historical data sample and perform another chi-square test. We keep candidates with a 5th to 95th chi-square statistic less than 5 in every filter. The 5th to 95th percentile test offers a measure of variability resistant to outliers, thus we can be more restrictive with the upper limit.

We seek to eliminate candidates with periodic variability and narrow our search to alerts with morphological differences from their respective archival light curves. The two-sample K-S test determines whether or not two different samples of data could have been drawn from the same distribution. We use a 95\% confidence interval to test if data from the historic ZTF light curves and alert-epoch light curves could have been drawn from the same distribution. We keep candidates with p-values less than 0.05. These candidates likely have alert data drawn from a different distribution than the historical data. 

\subsubsection{P-value Simulations and Restrictions on Dip Amplitude}

The 0.05 p-value threshold we imposed for the two-sample K-S test is unique to this workflow and dependent on ZTF schematics. We demonstrate a process of empirically determining such a threshold to inform future tests on early Rubin data.

We investigated how the p-value depends on the depth of the dip in brightness by simulating a range of dip depths for several candidates. Figure \ref{fig:sims} shows one such plot, where simulated data is plotted in color while real historical and alert data for the source, ZTF24abbldyz is plotted in grey. To simulate the dips, we generated 201 lists of random data with median offsets from historical data ranging from 0 to 0.2 magnitudes. Each set of random data has a length equal to the actual number of alert data points and a standard deviation equaling that of the historical data. We then performed a two-sample K-S test comparing the distribution of historical data to that of each simulated dip. In Figure \ref{fig:p-values} we plot p-value as a function of dip depth for ZTF24abbldyz. We replicated this plot for several different sources and found that the shape of the distribution was similar across all objects and filters. We conclude that a p-value cut-off of 0.05 seems reasonable as this cut-off ensures dips greater than about 0.01 magnitudes are considered statistically significant. Alerts with clear visible dips pass our K-S test with p-values well under the 0.05 threshold. This criterion will almost certainly need to be adjusted in future work to make robust cuts given the survey schematics and unique considerations of each anomaly search.

To impose a physical restriction on dip depth, we require that the median magnitude of the alert data sample be fainter than the 95th percentile of the historical data. We use the median as a measure of center because it is resistant to outliers in our samples. 

\subsubsection{Workflow Outputs}

After both rounds of filtering, we save light curves for the remaining candidates and append all relevant statistics and information to our growing database. We include a column with the number of times the object has appeared in the database so far, such that we can track how many times a single object has triggered an alert and passed all of our filter constraints. All other columns are described in our database schema, which can be found on \href{https://github.com/eleanorgallay/alert_seti}{GitHub}.

We also plot and save a color-magnitude diagram including all candidates that pass the initial Lasair filter on a given night. We pull data from the Gaia catalog of nearby stars to plot a 2D histogram color-magnitude diagram. We then perform a cone search in the Gaia database to find the closest source match for each object that passes the initial filter in order to highlight where our candidates fall on the diagram.
\begin{figure}[!ht]
    \centering
    \includegraphics[width=1\linewidth]{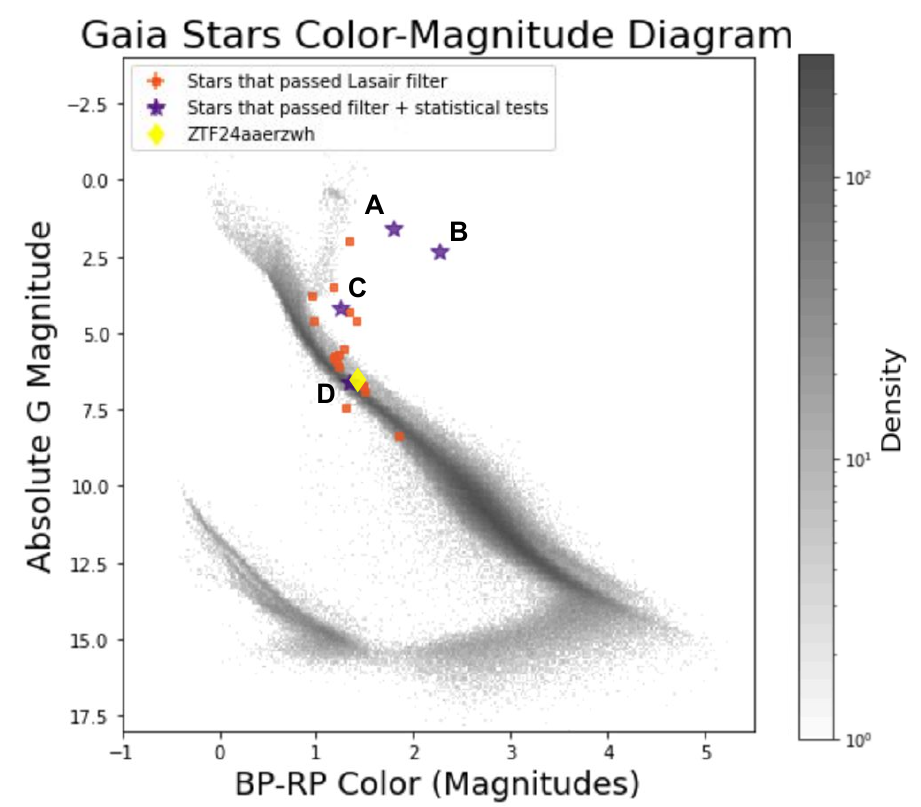}
    \caption{Color-Magnitude Diagram from 2024 November 25, showing alerts from the night of 2024 November 24. Alert data points are plotted in front of a 2D histogram color-magnitude diagram of the Gaia catalog of nearby stars. Objects that passed all filter constraints from the SQL query posed via the Lasair API are plotted as orange squares. Objects that passed both the initial Lasair filter and the second round of statistical tests and other constraints beyond the brokers are plotted as purple stars. The letters correspond to the ZTF object names associated with each purple star: A for ZTF24abbldyz, B for ZTF24abjinpg, C for ZTF24abgzfsz, and D for ZTF24ablvgno. Note that point D is partially obstructed by the yellow diamond. Although not from the night of 2024 November 24, we plot ZTF24aaerzwh, our curious long period variable, as a yellow diamond. This long period variable is clearly situated on the main sequence. }
    \label{fig:cmd}
\end{figure}

\begin{figure*}
    \centering
    \includegraphics[width=0.49\linewidth]{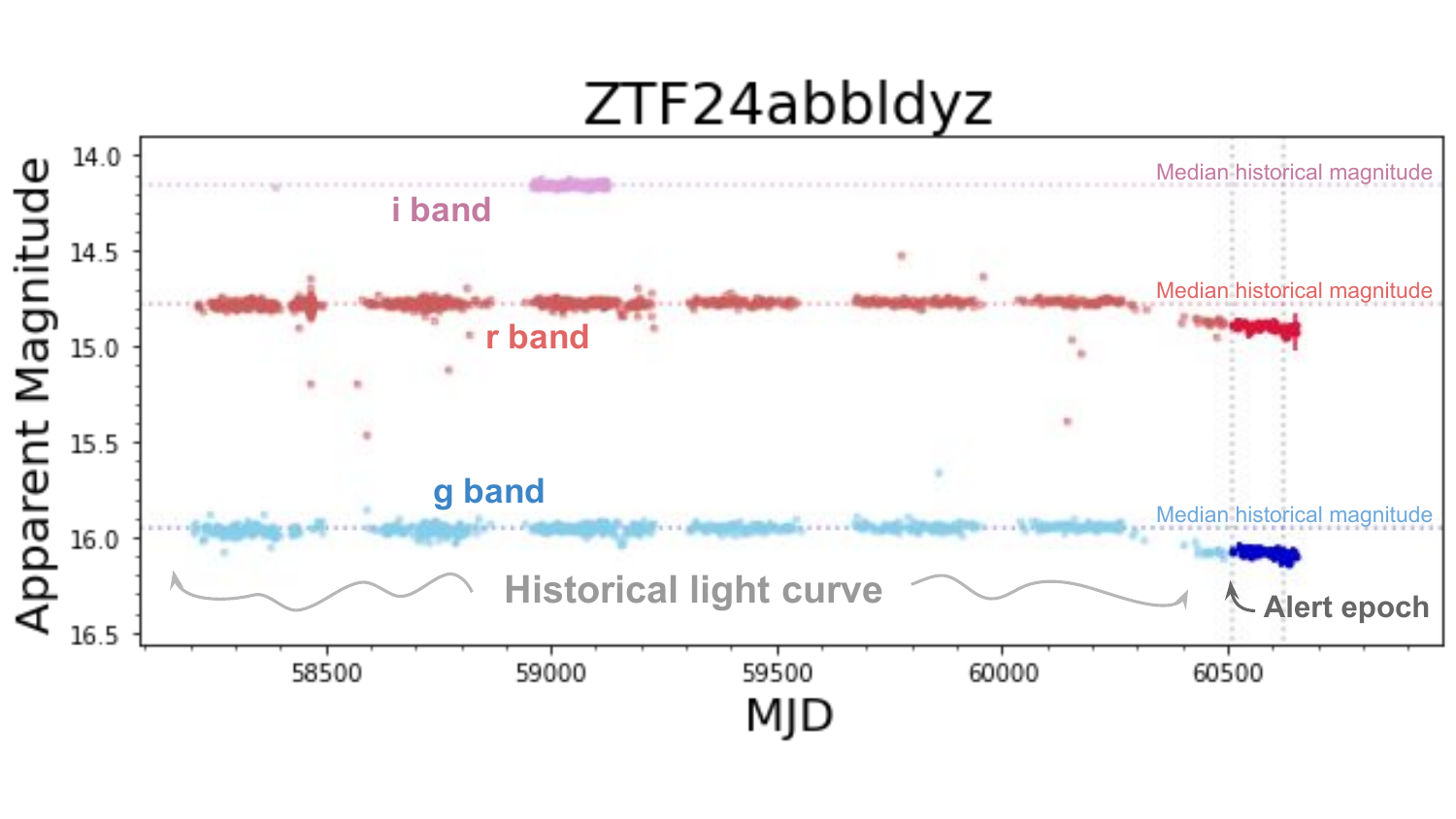}
    \includegraphics[width=0.49\linewidth]{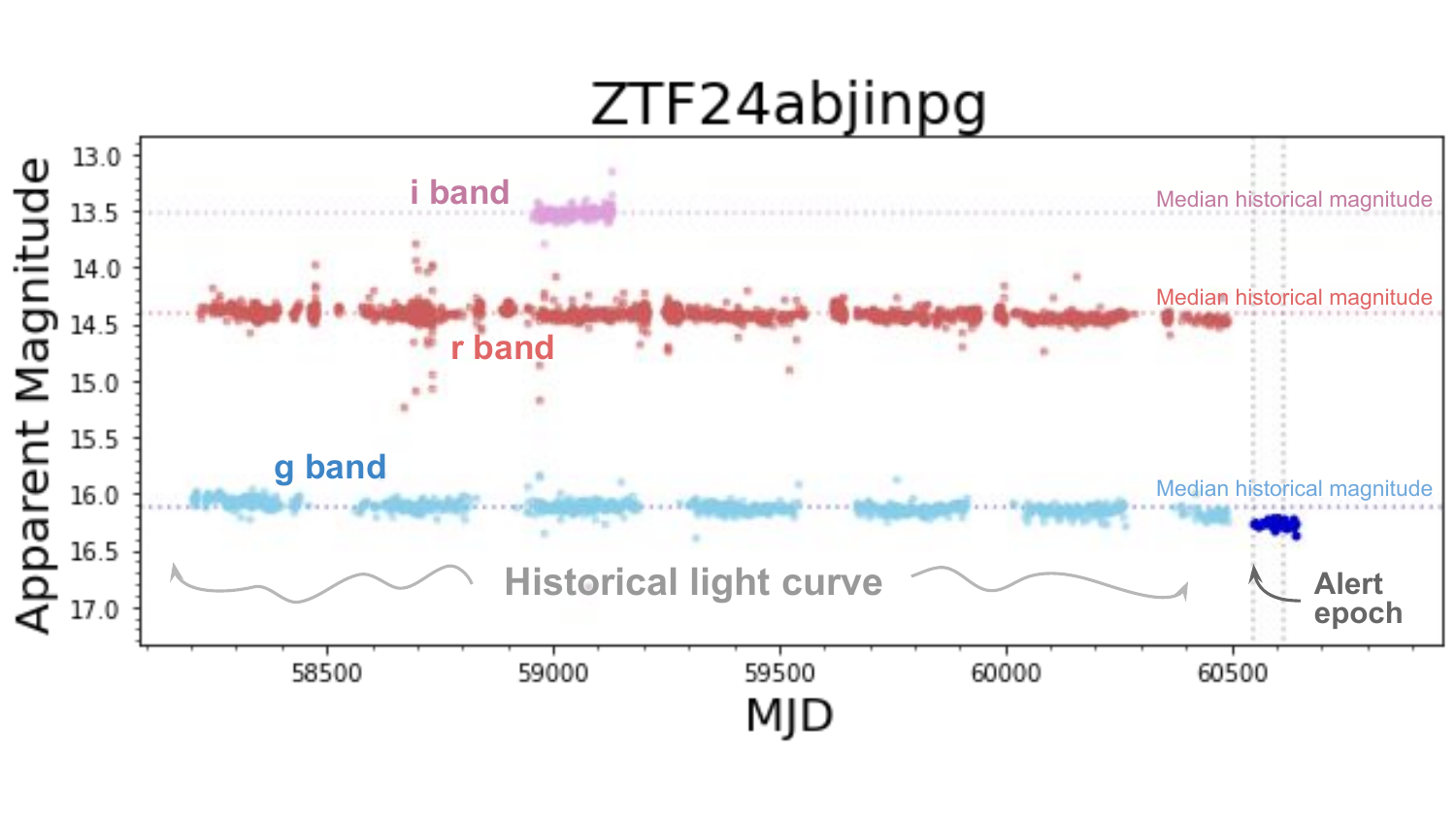}
    \includegraphics[width=0.49\linewidth]{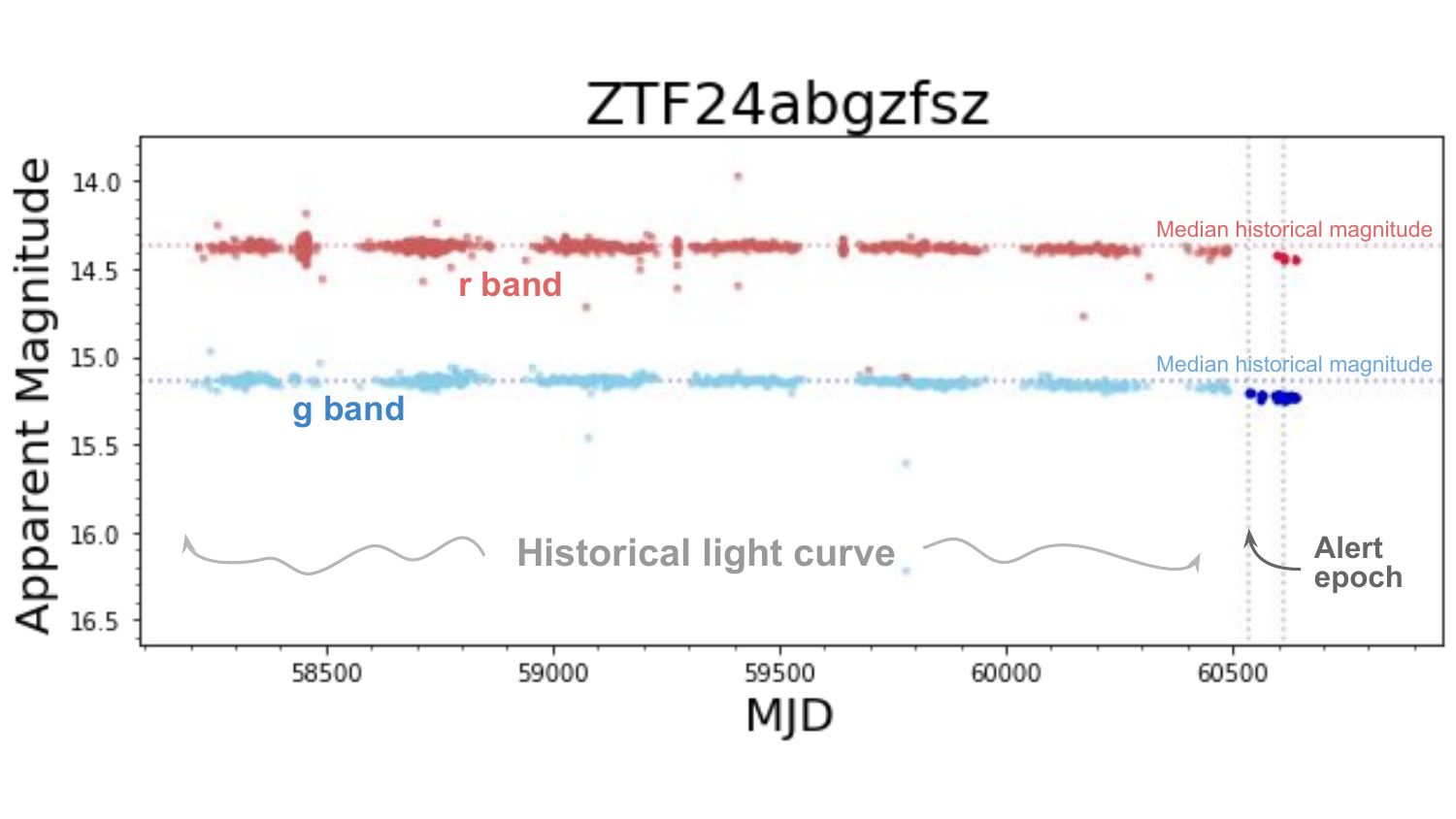}
    \includegraphics[width=0.49\linewidth]{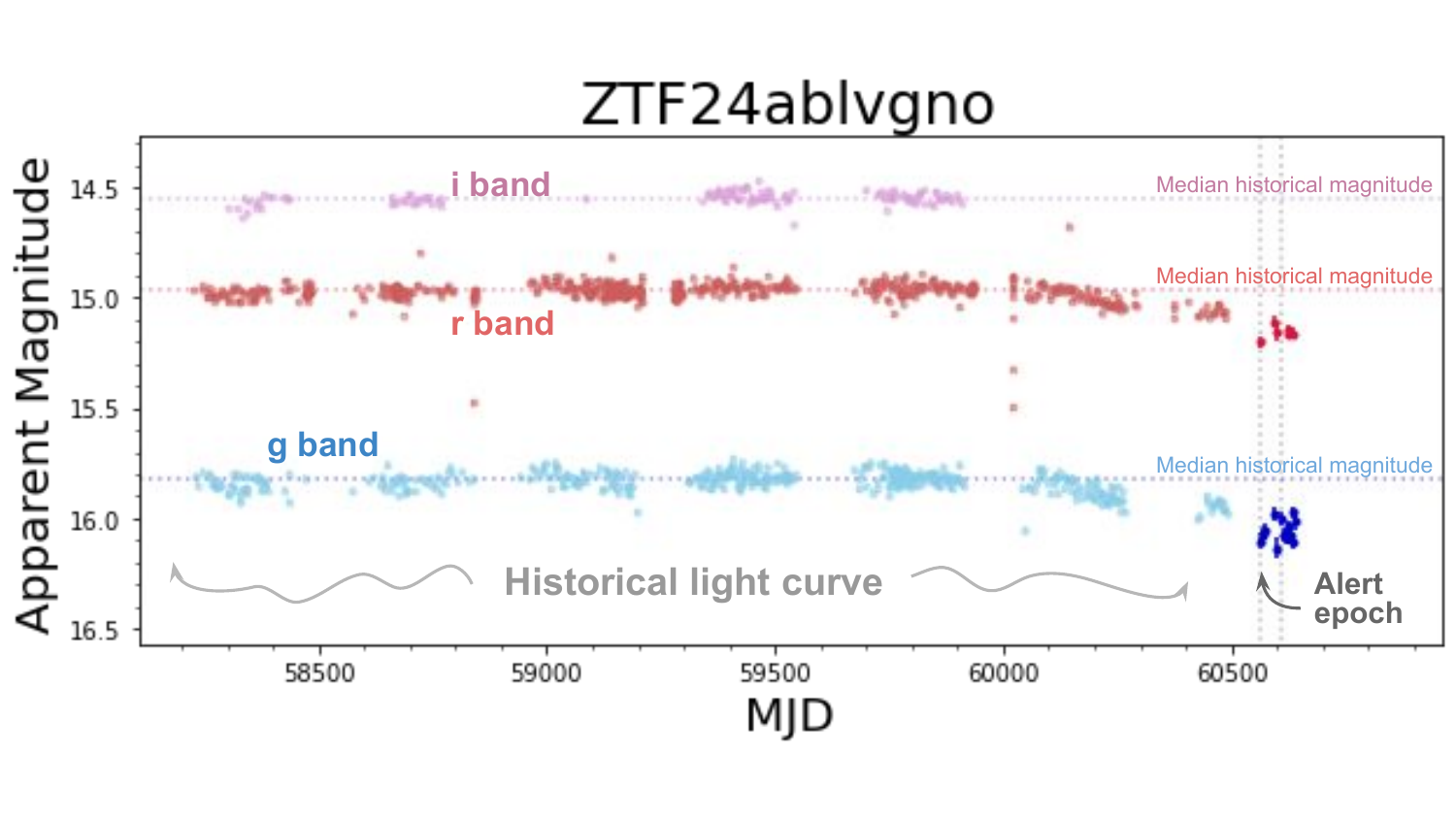}
    \caption{Light curves are plotted for all four objects that passed both sets of filters within and beyond the brokers on the night of 2024 November 24. Left to right and top to bottom, we have ZTF24abbldyz, ZTF24abjinpg, ZTF24abgzfsz, and ZTF24ablvgno. Figure \ref{fig:cmd} shows these objects plotted on a color-magnitude diagram as points A, B, C, and D, respectively. As in Figure \ref{fig:lpv}, the data plotted is an amalgamation of historical data pulled from the ZTF archive and alert packet data pulled with the Lasair API query. I-band, R-band, and G-band data are shown in purple, red, and blue, respectively. Note that I-band data is the least represented of the three bands, thus we primarily use R-band and G-band data to draw conclusions about the objects' behavior. All other plot features are the same as in Figure \ref{fig:lpv}. It can be observed in all four light curves that the alert packet data is fainter than the historical data, indicating a dip in brightness. ZTF24abbldyz shows a gradual dip in brightness beginning well before the alert epoch ($\approx$ 250 days earlier) and continuing until the most recent data. Despite having noisier data, ZTF24ablvgno shows a clear dip in brightness beginning roughly 500 days before the alert epoch.}
    \label{fig:lightcurves}
 \end{figure*}

\subsection{Example Results}

On a typical night, our initial filters within the Lasair broker narrow the ZTF stream of roughly one million alerts to under 20. After our second set of constraints, statistical tests, and restrictions on minimum historical sample size and dip depth, we tend to narrow our list of candidates to fewer than 5. 

The color-magnitude diagram from the night of 2024 November 24 is shown in Figure \ref{fig:cmd}, with one point added to show ZTF24aaerzwh, the long period variable that caught our attention on a different night of observation months earlier. On the night of 2024 November 24, 21 alerts flagged by ZTF met the criteria in our Lasair query, of which 4 passed our second set of filters. All 21 objects are located along the main sequence or slightly offset below the giant branch. We interpret the stars positioned below the giant branch as subgiants whose light is obscured by dust extinction. Note that our workflow also finds Young Stellar Objects (YSOs), though none were found on this particular night of observation. 

Further development of the workflow could include a study of where all candidates lie on the color-magnitude diagram. The diagram situates anomalous dippers within the context of stellar evolution. The magnitude of a dip may have different significance depending on whether the star lies on the main sequence or has progressed into a later stage of evolution. For instance, one might consider a dimming event more or less anomalous depending on where it is situated on the diagram. Comparing high vs. low mass stellar dippers may offer additional insight into the dimming events' sources and significance, thus future optical technosignature searches may benefit from the information in the color-magnitude diagram.

Typical light curves we find using this workflow show relatively quiet historical data with a slight dip in magnitude beginning at or just before the alert epoch. Dip amplitudes are visually apparent though rarely greater than about a quarter magnitude. Four light curves from the night of 2024 November 24 are shown in Figure \ref{fig:lightcurves}. All four objects passed the initial Lasair filter and additional statistical tests. The stars each begin to dip in brightness before the alert epoch, suggesting their dip amplitudes did not meet ZTF's alert threshold prior to their discovery dates. The minimum dip amplitude required to trigger an alert in ZTF remains unclear, and further research could explore the alert criteria, as this information is not publicly available at the time of writing. 

The light curves resulting from our workflow indicate we have successfully identified new stellar dippers using real time alerts. To verify these are bona fide dippers, or technosignature candidates, additional analysis of the pixel-level data should be done to rule out systematic contamination or calibration problems. This could include analyzing nearby stars for similar behavior over time, as was done in the case of Boyajian's Star \citep{boyajian2016}. Although detailed follow up on these candidates is beyond the scope of this paper, our initial results make clear the promising opportunity for future anomaly and technosignature searches using alert brokers on data from ZTF and the upcoming LSST on the Vera C. Rubin Observatory.

\section{Discussion}

Alert brokers offer tools to help pick out anomalous signals that might suggest life beyond Earth. We have investigated methods of conducting technosignature searches using the current broker technology. Implementing the unique watchmap feature offered by the Lasair alert broker, we find some SETI projects are possible using features directly in the brokers such as searches for anomalous signals in the transit zones of solar system planets. Other projects with temporal constraints, like SETI Ellipsoid and searches for novel variability, are not possible relying solely on current broker technology and may require additional programming and manual steps. 

We demonstrate a specific example of such a project using current broker technology together with methods extending beyond broker capabilities, showing a fully fleshed-out workflow to find new stellar dippers. The first stage of our process involves filtering within the Lasair broker, either by creating a SQL filter on their website or by replicating the SQL conditions as an API query. This first filter narrows the data stream from ZTF from 1 million alerts to around 20 or fewer each night. The second stage of the workflow is not currently possible within the brokers, and involves comparing alert data with historical data pulled from the ZTF archive to remove alerts that are caused by predictable rather than anomalous variability. After imposing additional filters and constraints outside the broker, we pair down our list to around 5 or fewer candidates for possible follow-up. 
This search method is designed to return anomalous stars which may have SETI implications, though further observation and analysis are required to distinguish technosignatures from natural astrophysical anomalies.
We emphasize that this demonstration has used simple statistical techniques, such as K-S tests and $\chi^2$ thresholds, that are appropriate to the current ZTF archive. However, many other statistical methods for detecting outlier or anomalous light curve behavior, including machine learning approaches such as Random Forest clustering and light curve similarity searches \citep[e.g.][]{aleo2024}. These other techniques will be especially important for early analysis of the LSST data, which will not have the multi-year archive that we use from ZTF for baseline comparison.

With the upcoming LSST, which is expected to increase the alert stream by a factor of 10, we anticipate our workflow will turn up roughly 50 candidates every night, increasing our likelihood of finding a technosignature by an order of magnitude. The search methods discussed in this paper work on all alert surveys and older instruments like ZTF, though LSST is expected to be the primary source of alert data for the next decade. Situated in the southern hemisphere, LSST will bring in much new data for stars that have not yet been studied by previous alert surveys, mainly located in the north. With the large volume of alerts coming in from LSST every night, the datastream will be intractable to search locally. Alert brokers enable efficient searches for anomalous stars and technosignatures among the millions of nightly alerts that LSST will flag in the coming years.

Limitations to current broker technology lie mainly in the information that is not visible to users. The brokers could be enhanced by providing access to apparent magnitudes, object identifications, and archival data, which would facilitate spatial and temporal technosignature searches as well as broader anomaly detection. We have shown however that existing broker capabilities can be adapted to conduct anomaly searches in real-time. Brokers offer tools for quick identification of technosignature candidates, which may be critical for follow-up. While tedious, technosignature search methods within and beyond the broker framework show strong potential for early science with LSST. \\


The authors thank the anonymous referee whose suggestions greatly improved the quality of this manuscript. 
The authors wish to thank Chris Lintott, Andrew Siemion, and Anastasios Tzanidakis for productive conversations that helped shape the work, and David MacMahon for help with the transit zone geometry.

JRAD and SC acknowledge support from Breakthrough Listen. The Breakthrough Prize Foundation funds the Breakthrough Initiatives, which manages Breakthrough Listen. EG was funded as a participant in the Berkeley SETI Research Center Research Experience for Undergraduates Site, supported by the National Science Foundation under Grant No.~2244242.

JRAD acknowledges support from the DiRAC Institute in the Department of Astronomy at the University of Washington. The DiRAC Institute is supported through generous gifts from the Charles and Lisa Simonyi Fund for Arts and Sciences, Janet and Lloyd Frink, and the Washington Research Foundation. \\


\software{
Python, IPython \citep{ipython}, 
NumPy \citep{numpy}, 
Matplotlib \citep{matplotlib}, 
SciPy \citep{scipy}, 
Pandas \citep{pandas}, 
Astropy \citep{astropy:2013,astropy:2018, astropy:2022},
Astroquery \citep{ginsburg_astroquery_2019},
ZTFquery \citep{rigault_ztfquery_2018},
MOCpy \citep{fernique_moc_2014}
}

\bibliography{eleanoreferences,references}



\appendix
\section{SQL Code}
\label{ap:SQL}

For reference, here we include the full SQL query used with  the Lasair API in Python to search for new stellar dippers. We define the current date MJD and JD as \texttt{mjdnow} and \texttt{jdnow}, respectively, and define a variable called \texttt{days} in order to specify the window of time in which to pull search results. Here we select novel dippers in the past 24 hours. We select all columns from the objects and sherlock\_classifications tables. Each line in the conditions section is a filter constraint. For instance, we require that the object has a gaia crossmatch (condition 7) and no alert-packet data points fainter than its reference magnitude (condition 11). The full query is:

\lstset{
    language=Python,
    basicstyle=\ttfamily\footnotesize, 
    keywordstyle=\color{blue},          
    commentstyle=\color{gray},          
    numberstyle=\tiny\color{gray},      
    stepnumber=1,                       
    numbersep=5pt,                      
    backgroundcolor=\color{white},  
    showspaces=false,                   
    showstringspaces=false              
}
\begin{lstlisting}
    
mjdnow = str(Time.now().mjd)
jdnow = str(Time.now().jd)
days = str(1) # number of days 

selected = '*'

tables = 'objects,sherlock_classifications'

conditions = """
objects.objectId=sherlock_classifications.objectId
AND (objects.sgscore1 > 0.9)
AND (sherlock_classifications.classification != "SN")
AND (sherlock_classifications.classification != "NT")
AND (sherlock_classifications.classification != "AGN")
AND (objects.ncand >= 10)
AND (sherlock_classifications.catalogue_table_name LIKE "%gaia%")
AND (objects.objectId LIKE "ZTF24%")
AND sherlock_classifications.separationArcsec < 0.5
AND ((objects.sgmag1 < 16)
   OR (objects.srmag1 < 16))
AND ISNULL(objects.ncandgp)
AND ("""+jdnow+"""- objects.jdmax) < """+days+"""

"""

L = lasair(settings.API_TOKEN, endpoint = "https://lasair-ztf.lsst.ac.uk/api")

try:
    v4 = L.query(selected, tables, conditions)
except LasairError as e:
    print(e)

\end{lstlisting}

\end{document}